\documentclass{nature}
\usepackage{graphicx}
\usepackage{color}
\usepackage{times}
\usepackage{graphicx}
\usepackage{dcolumn}
\usepackage{bm}
\usepackage{amssymb,amsmath}
\usepackage{verbatim}
\usepackage{subfigure}
\usepackage{textcomp}
\usepackage[utf8]{inputenc}
\usepackage[T1]{fontenc}
\usepackage{multirow}
\usepackage{booktabs}
\usepackage{color}

\title{Evanescent single-molecule biosensing with quantum limited precision}
\author{N. P. Mauranyapin$^1$, L. S. Madsen$^{1,2}$, M. A. Taylor$^{1,3}$, M. Waleed$^1$ \& W. P. Bowen$^{1,2,*}$}

\begin{document}

\maketitle

\begin{affiliations}
 \item School of Mathematics and Physics, the University of Queensland, Australia
 \item Centre for Engineered Quantum Systems, the University of Queensland, Australia 
 \item Research Institute of Molecular Pathology, Vienna, Austria.
\end{affiliations}

\abstract{
Sensors that are able to detect and track single unlabelled biomolecules are an important tool both to understand biomolecular dynamics and interactions at nanoscale, and for medical diagnostics operating at their ultimate detection limits. Recently, exceptional sensitivity has been achieved using the strongly enhanced evanescent fields provided by optical microcavities and  nano-sized plasmonic resonators. However, at high field intensities photodamage to the biological specimen becomes increasingly problematic. Here, we introduce an optical nanofibre based evanescent biosensor that operates at the fundamental precision limit  introduced by quantisation of light. This allows a four order-of-magnitude reduction in optical intensity whilst maintaining state-of-the-art sensitivity. It enables quantum noise limited tracking of single biomolecules as small as 3.5~nm, and surface-molecule interactions to be monitored over extended periods.
By achieving quantum noise limited precision, our approach provides a pathway towards quantum-enhanced single-molecule biosensors.

}
\rm

\section*{Introduction}

Evanescent optical biosensors that operate label-free and can resolve single molecules have applications ranging from clinical diagnostics\cite{dantham}, to environmental monitoring\cite{zhu, li} and the detection and manipulation of viruses\cite{keng}, proteins and antibodies\cite{ baaske, pang, zijlstra}. Further, they offer the prospect to provide new insights into motor molecule dynamics and biophysically important conformational changes as they occur in the natural state, unmodified by the presence of fluorescent markers or nanoparticle labels\cite{baaske}. Recently,  the reach of evanescent techniques has been extended to single proteins with Stokes radii of a few nanometers\cite{pang,dantham} by concentrating the optical field using resonant structures such as optical microcavities\cite{baaske, keng, zhu} and plasmonic resonators\cite{pang,dantham,zijlstra}. These advances illustrate a near-universal feature of precision optical biosensors --- that increased light intensities are required to detect smaller molecules or improve spatiotemporal resolution. This increases the photodamage experienced by the specimen, which can have broad consequences on viability\cite{mirsaidov}, function\cite{sowa}, structure\cite{landry} and growth\cite{waldchen}. It is therefore desirable to develop alternative biosensing approaches that improve sensitivity without 
exposing the specimen to higher optical intensities. 

Here we demonstrate an optical nanofibre-based approach to evanescent detection and tracking of unlabelled bio-molecules that utilises a combination of heterodyne interferometry and dark field illumination. This greatly suppresses technical noise due to background scatter, vibrations and laser fluctuations that has limited previous experiments~\cite{knittel,lu}, allowing operation at the quantum noise limit to sensitivity introduced by the quantisation of light. The increased  information that is extracted per scattered photon enables state-of-the-art sensitivity to be achieved with optical intensities four orders of magnitude lower than has been possible previously~\cite{dantham, pang}. 
Using the biosensor, we detect nanospheres and biomolecules as small as 3.5 nm in radius and track them with 5 nm resolution at 100 Hz bandwidth.

\vspace{5pc}

\section*{Dark-field nanofibre based biosensor}

\begin{figure}
\centering
\includegraphics[width=0.6\textwidth,height=0.8\textwidth,clip,keepaspectratio]{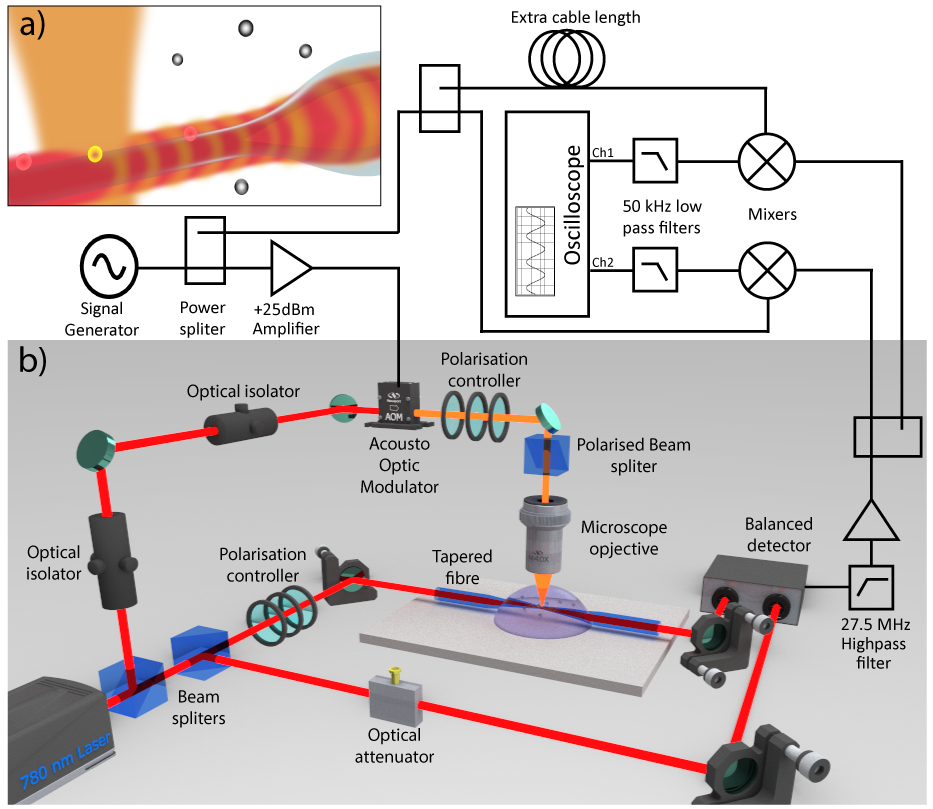}
\caption{\textbf{Experimental setup.}
\textbf{(a)} Nanofiber with dark field heterodyne illumination. 
Nanoparticles (grey spheres) in a droplet of ultra pure water are detected when entering the probe beam waist next to the nanofibre. 
 \textbf{(b)} Schematic of the optical setup for the dark field heterodyne nanofibre biosensor, including two optical isolators to suppress back-scatter of probe light which was found to increase the noise floor even for very low photon flux, a low noise New Focus 1807 balanced photoreceiver with electronic noise well beneath the local oscillator shot noise level, and a home-built ultralow noise dual quadrature electronic lock-in amplifier (methods).
}
\label{fig:setup}
\end{figure}

The nanofibre sensor is immersed in a droplet of water containing nanoparticles or biomolecules (see Fig.~\ref{fig:setup}a). Light passing through it induces an intense evanescent field extending a few hundred nanometers from the fibre surface. Likewise, light scattered in the nearfield of the nanofibre is collected by the guided mode of the fibre, providing a highly localised objective. In contrast to previous approaches\cite{swaim,yu}, we illuminate a small section of the nanofibre from above with a probe field. Since its propagation direction is orthogonal to the fibre axis, very little of the probe field is collected in the absence of nanoparticles and biomolecules. In this dark-field configuration, minimal background noise is introduced by the probe. When a nanoparticle or biomolecule diffuses into the illuminated region,  it scatters probe light by elastic dipole scattering. The nanofibre collects a significant fraction of this field, which we term the {\it signal field} henceforth. Diffusion of the particle in the vicinity of the nanofibre modulates the collection efficiency, encoding information about the motion of the particle on the signal field intensity.

\vspace{5pc}
\section*{Noise performance}

 Quantum noise limited tracking of single unlabelled biomolecules is made difficult by the combination of very low levels of scattered power --- in our case in the range of femtowatts  --- and technical noise sources such as laser intensity and frequency fluctuations, electronic noise, acoustic vibrations and background scattered light. These technical noise sources are particularly problematic in the hertz to kilohertz frequency band of relevance for observations of biomolecule dynamics, binding and trapping~\cite{taylor_janousek}, and are a key limitation of previously reported evanescent sensors~\cite{knittel,lu}.
To achieve quantum noise limited performance here, we use an optical heterodyne technique to amplify the signal from the trapped particle above both the electronic noise of our measurement system and noise from background light, and to shift its frequency well away from low frequency laser, electronic and acoustic noise.
In short, an optical local oscillator field frequency shifted by 72.58 MHz from the probe is injected into the nanofibre, and its beat with the signal field is observed on a low noise photoreceiver (see methods). A photocurrent proportional to the absolute value of the signal field amplitude is acquired in real time by mixing the photoreceiver output down at the heterodyne beat frequency using a home-built dual-phase lock-in amplifier (see Fig.~\ref{fig:setup}).  

\begin{figure}
\centering
\includegraphics[width=0.95\textwidth,height=0.8\textwidth,clip,keepaspectratio]{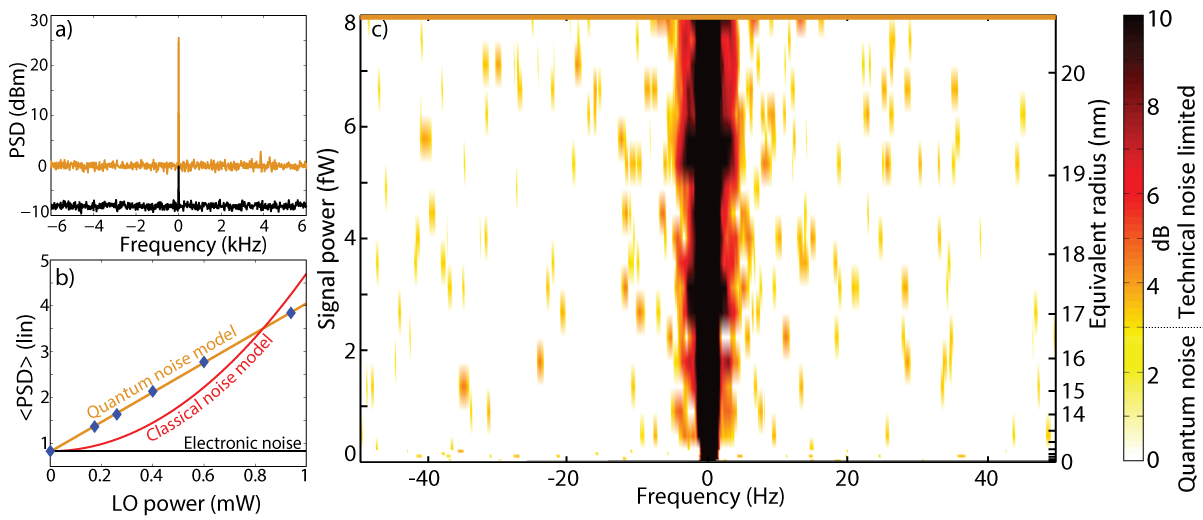}
\caption{\textbf{Quantum noise limited region}. 
\textbf{(a)} The power spectrum of the electronic noise and the optical background response of the system under normal operating conditions are represented by the black and orange curve, respectively.
\textbf{(b)} Averaged noise power spectrum over a 10~kHz bandwidth without probe light as function of the local oscillator power (blue points). The linear orange, quadratic red and constant black curves represent the quantum model, a classical laser noise model and an electronic noise model, respectively.
\textbf{(c)} Clearance from quantum noise floor as function of signal power (left y-axis) or its equivalent nanoparticle radius (right y-axis) and as function of frequencies (x-axis), for a stationary scattering source.
The signal power is estimated from the amplification of the scattered probe field by the local oscillator field (see methods). 
The noise floor is dominated by quantum noise in the white-yellow regions, and by technical noise in the orange-black regions. 
}
\label{fig:Q_lim}
\end{figure}

We performed a sequence of experiments to characterise the noise performance of the biosensor and verify its quantum-limited performance (Fig~\ref{fig:Q_lim}). These confirm that electronic dark noise can be neglected at frequencies above a few hertz (Fig~\ref{fig:Q_lim}a), and that the local oscillator field is quantum noise limited (Fig~\ref{fig:Q_lim}b). The total noise floor of the biosensor was measured as a static scattering center was illuminated with a range of powers, with results displayed in Fig.~\ref{fig:Q_lim}c as a function of collected signal power and equivalent particle size at fixed intensity. Quantum noise is found to dominate at all measurement frequencies above 4~Hz, even for the highest signal power used (8~fW, equivalent radius of 20.5~nm). 
This hertz-kilohertz frequency window is important  for biophysical processes ranging from seconds to few milliseconds~\cite{kitamura}, and is the crucial window for measurements of the motion of trapped nanoparticles and biomolecules as performed here.

The quantum noise limit of our biosensor can be quantified by comparing the shot noise level due to the quantisation of light to the scattered intensity predicted from Rayleigh scattering theory (Supplementary Information II.C.1).
The minimum detectable particle cross section is $\sigma_{min} = {8 \hbar \omega}/{\eta \tau I_{\rm probe} }$,
where $ \hbar $ is the reduced Planck constant, $ \omega$ the frequency of the probe light, $ \eta$ the total collection efficiency including detection inefficiencies, $ \tau$ the measurement time and $ I_{\rm probe} $ the probe intensity. 
We estimate the collection efficiency to be in the range of 1--10\%, with more accurate determination precluded by a strong dependence on both the nanofibre geometry and the particle position.
For our experimental parameters of $I_{\rm probe} = 7 \times 10^7 $Wm$^{-2} $ and $ \tau = 10$~ms,  the smallest detectable particle cross section is then predicted to be in the range of $ 3 \times 10^{-5} $  to $ 3 \times 10^{-4}$~nm$^2$, corresponding 
to a silica nanosphere of radius 16--23~nm.

\vspace{5pc}
\section*{Detection and trapping of single nanoparticles}

\begin{figure}
\centering
\includegraphics[width=0.8\textwidth,height=0.6\textwidth,clip,keepaspectratio]{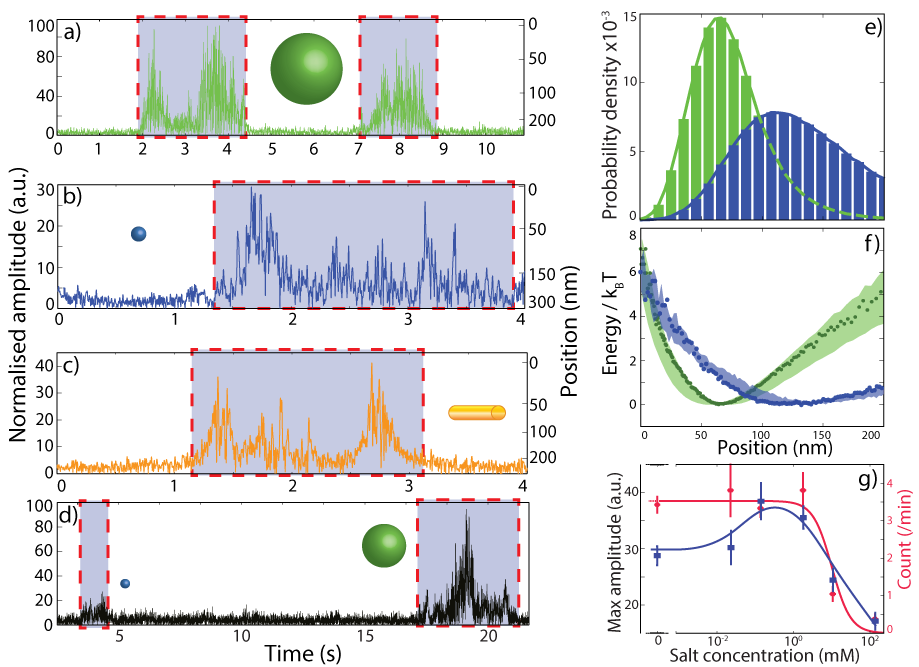}
\caption{\textbf{ Nano particle detection and trapping}. \textbf{(a)}, \textbf{(b)} and \textbf{(c)} Time trace of the normalised amplitude (left y-axis) and corresponding position (right y-axis) of the detected signal in which a sudden rise (shaded regions) indicates trapping events of 25~nm, 5~nm silica sphere and 10 by 45~nm gold nanorod. For clarity, these time traces are bandpass filtered over the range 4 to 100~Hz.
\textbf{(d)} Time trace of the normalised amplitude for a solution containing both 25~nm and 5~nm silica particles.
\textbf{(e)} Mean probability density associated with all observed trapping events of the 25~nm (11 events) and 5~nm (8 events) particles.
\textbf{(f)} Calculated trapping potentials 
derived from one trapping event for 25~nm (green) and 5~nm (blue) particles. The shaded bands represent the standard deviation of the potential for all observed 5~nm and 25~nm trapping events. 
\textbf{(g)} Two independent experiments monitoring the maximum amplitude (blue, left axis) of the trapping events and their frequency (red, right y-axis) as a function of salt concentration for 5 nm silica nanospheres. The curves are guides-to-the-eye  and the error bars represent the standard error of each measurement. The total number of detection events for salt concentration of \{$\sim 0$, $5.5 \times 10 ^{-2}$, $2.8 \times 10 ^{-1}$, $2.8$, $14$, $1.4\times 10 ^{2}$\} mM, were \{25, 26, 24, 27, 8, 3\} and \{9, 10, 9, 11, 6, 2\} for the count rate and amplitude experiment respectively.
}
\label{fig:NP}
\end{figure}

The biosensor was tested on solutions of silica and gold nanoparticles in 
ultrapure double processed deionized water (Sigma W3500). 
Unexpectedly, the sensor is able to resolve silica nanoparticles of radius down to 5~nm,  significantly beneath the quantum noise limit  calculated in the previous section. We attribute this to an enhanced scattering cross-section due to the presence of surface charges on the nanoparticles, as later discussed. 
Figures~\ref{fig:NP}a-c show sections of typical time domain traces that display nanoparticle detection events for 25~nm and 5~nm silica particles, and 10$\times$45~nm gold nanorods, respectively. 
Calibration of the detected signal in terms of the particle position reveals that the 5 and 25 nm nanoparticles can be tracked with resolution of 5 and 1 nm, respectively, with a 100 Hz bandwidth  (details on calibration in Supplementary Information VII).

Further experiments were performed using a solution containing both 5~and~25~nm nanoparticles (Fig.~\ref{fig:NP}d).  The observed 
events are similar to those for individual particle solutions (Fig.~\ref{fig:NP}a,b)
showing that parallel detection and discrimination of different nanoparticle types is possible.
These results compare favourably to other nanofiber sensors, with the smallest nanoparticles previously observed having a 100 nm radius~\cite{yu}. Moreover, the system is competitive with the best field-enhanced evanescent sensors both using micro-cavities\cite{baaske, dantham, keng, zhu} and plasmonic resonators\cite{pang,zijlstra}, while exposing the specimen to significantly lower optical intensities (see Discussion). This demonstrates the substantial performance gains that can be achieved via dark field heterodyne detection and complements the recent demonstration of quantum noise limited super-resolution imaging of stationary proteins~\cite{piliarik}.

The power spectrum of the measured particle motion exhibits a Lorenzian shape characteristic of trapped Brownian motion (Supplementary Information V), which confirms that the particles are trapped.
By calculating the mean position probability distributions obtained from the 5~and~25~nm nanospheres events (see Fig.~\ref{fig:NP}e) and using Boltzmann statistics, we retrieve the trapping potentials shown in Fig.~\ref{fig:NP}f (Supplementary Information VI). 
Unlike previous experiments\cite{swaim,yu} which used a combination of repulsive electrostatic forces and attractive optical forces to trap larger particle near the nanofibre, we demonstrate theoretically and experimentally that optical attraction is negligible for our smaller particles (Supplementary Information VIII).

Recently, the surface charges that accumulate on surfaces in solution and their associated counter-ions have been shown to greatly enhance the scattering cross section of micelles smaller than 200 nm trapped in an optical tweezer\cite{park}. Since the surface area-to-volume ratio increases with reduced particle size, a considerably stronger effect could be anticipated in experiments with nanoparticles \cite{stoylov}. Surface charges have also been shown to lead to long range attractive forces between same-charged particles~Ref.\cite{nagornyak}.
We believe that this combination of effects, due to the surface charges present on both nanofibre and nanoparticles, may explain the enhanced scattering cross section and the ability to trap in our experiments. 
We rule out alternative explanations such as outside contaminants and aggregation based on the reproducibility of the signal amplitudes and potentials and control experiments (Supplementary Information IX). 
To test the feasibility of signal enhancement via surface charges, 
we repeated the experiment with 5 nm particles while varying the salt concentration in the solution using DPBS (Dulbecco’s Phosphate Buffered Saline, Gibco 14040). The spatial extent of the counter-ion distribution generated by surface charges varies strongly with salt concentration due to screening effects. In ultrapure water, as in our initial experiments, it can extend over several hundreds nanometers and therefore significantly influence experiments, while it is reduced to nanometer-scale for even relatively modest salt concentrations\cite{larsen}.
As shown Fig.~\ref{fig:NP}g both the amplitude of light scattered by the nanoparticles and the frequency of events are effected by salt concentration, dropping significantly above a concentration of around 1~mM, similarly to the observations in Ref.~\cite{park} and consistent with a decrease in scattering cross section and attractive forces.
Further studies are required to provide a
more definitive explanation of these effects, to test whether they might explain previous observations of enhanced sensitivity in evanescent biosensors\cite{armani}, and to explore their wider potential as an enhancement mechanism to improve detection limits of those sensors.

\vspace{5pc}
\section*{Detection and trapping of single unlabelled biomolecules}

\begin{figure}
\centering
\includegraphics[width=1\textwidth,height=0.8\textwidth,clip,keepaspectratio]{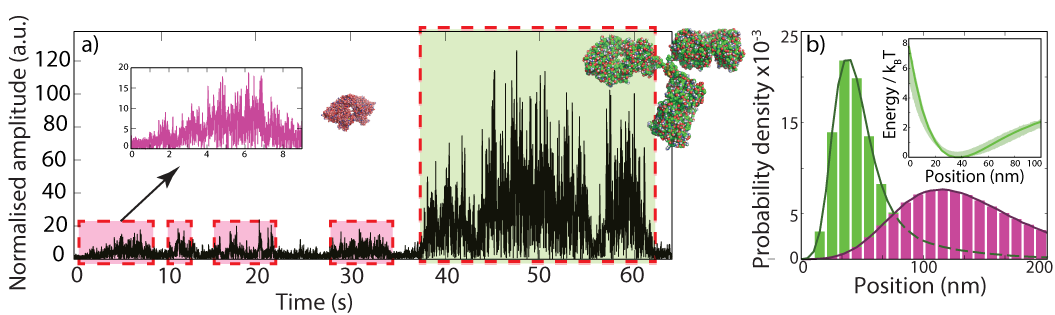}
\caption{\textbf{Biomolecule detection and trapping}. 
\textbf{(a)} Time trace of the normalised amplitude for a solution containing both BSA (0.09~mg/mL) and anti-{\it E. coli} antibody (0.03~mg/mL). The magenta regions represent trapping events of single BSA molecules and the green region an anti-{\it E. coli} antibody. 
\textbf{(b)} Mean probability distribution obtained from all observed trapping events of BSA and anti-{\it E. coli} antibody, shown in magenta and green respectively. Distributions calculated from a total of 11 and 13 events for anti-{\it E. coli} antibody and BSA, respectively. The insert shows the trapping potential derived from the anti-{\it E. coli} antibody probability distribution, and the shaded band its standard deviation.
}
\label{fig:bio_mol}
\end{figure}

With both quantum noise limited performance and the capability to detect small nanoparticles confirmed, we now apply our biosensor to the detection of single unlabelled biomolecules, an application that has not previously been demonstrated using a nanofibre sensor. We perform measurements on low concentration solutions of the biomolecules bovine serum albumin (BSA) and anti-{\it Escherichia coli} ({\it E. coli}) antibody, with molecular weights of 66~kDa and 150~kDa, respectively. BSA, in particular, has a 3.5 nm Stokes radius, and is among the smallest biomolecules detected using plasmonic and cavity-enhanced techniques\cite{pang,dantham}.
After performing a series of experiments using solutions of each biomolecule individually (not shown here), we demonstrate the ability to perform measurements in parallel using a solution containing both molecules. A time-trace which exhibits a sequence of trapping events of these molecules is shown in Fig.~\ref{fig:bio_mol}a. A factor of five difference between the amplitudes of anti-{\it E. coli} antibody and BSA events allows straightforward discrimination between the molecules. This ability to discriminate is also seen in the position probability distributions of the biomolecules, shown in Fig.~\ref{fig:bio_mol}b.
Using the same methods as for nanoparticles, we calculate the trapping potential from the probability distribution as shown in Fig.~\ref{fig:bio_mol}b~(inset) for an anti-{\it E. coli} antibody. The repulsive part of the potential provides information about the interaction between the molecule and the fibre surface, and could be used to monitor surface particle interactions~\cite{lopez}, cell membrane formation~\cite{sun}, molecule-molecule interactions~\cite{baaske} and molecular motor motion~\cite{fazal} with high sensitivity, in real time, and without recourse to ensemble averaging.

The ability to detect 3.5~nm biomolcules demonstrates that the performance of our biosensor is competitive in sensitivity with more complicated whispering gallery resonators and plasmonic evanescent sensors~\cite{pang,dantham},  though here this is achieved with four orders of magnitude lower light intensity (Supplementary Information X) 
and a considerably simpler, robust, sensing platform.

\vspace{5pc}
\section*{Discussion}

 Specimen photodamage due to exposure to light is often a critical issue in biophysical experiments~\cite{mirsaidov}, resulting in photochemical changes to biological processes~\cite{mirsaidov,sowa}, modifying structure and growth~\cite{landry,waldchen}, and ultimately adversely affecting viability~\cite{mirsaidov,waldchen}. For instance, Ref.~\cite{mirsaidov} found that the viability of  {\it E. Coli} is affected by light intensities as low as $4.9 \times 10^9$~Wm$^{-2}$. The experiments reported here used probe and local oscillator field intensities of $7 \times 10^7$~Wm$^{-2}$ and $7 \times 10^8$~Wm$^{-2}$, below this threshold; while similarly-performing plasmonic sensors use intensities of  around $10^{13}$~Wm$^{-2}$~\cite{pang}. More generally, the four order of magnitude reduction in intensity afforded by our biosensor 
 should allow a commensurate increase in observation time for equivalent photodamage\cite{mirsaidov}.
 
The combination of high sensitivity and bandwidth with low photodamage opens up a new path to explore single molecule biophysics. For example, it may allow single discreet steps in a free flagella motor to be observed at their natural frequencies. To date, such  steps have only been observed by attaching a fluorescent label to the flagella and slowing down the rotation of the motor by inducing photodamage to reduce the sodium-motive force\cite{sowa}. The bandwidth of the sensor can in principle be extended to the limits of our optical detector and digitalisation device, with bandwidths above 10 MHz feasible. This could allow studies of small scale conformational changes, such as occur in protein side chain and nucleic acid base conformation, occurring on time-scales of $10^{-11}$s to $10^{-6}$s ~\cite{mayor}, in solution, without markers and with minimal photodamage.

Since our biosensor reaches the quantum noise limit, it can be combined with quantum correlated photons to achieve sub-shot noise limited precision. 
 Numerous approaches to quantum enhanced sensing have been developed over the past few decades~\cite{taylor_bowen}. However, as yet they have not been applied to single molecule sensing. Finally, at the cost of increased intensity, our biosensing approach could be combined with plasmonic sensing by depositing plasmonic particles on the nanotaper. In this way, it may be possible to achieve quantum noise limited plasmonic sensing and thereby allow the detection of even smaller molecules than is currently possible.

\vspace{5pc}

\begin{methods}

\subsection{Heterodyne concept:}
The heterodyne strategy is similar to the approach developed in Refs. \cite{taylor_janousek,taylor_knittel} to evade low frequency noise sources in optical tweezers based biological measurements but, by using heterodyne detection, eliminates the need to phase stabilise the signal field.  To implement the technique, an optical local oscillator field is injected into the nanofibre, frequency shifted from the probe field by 72.58~MHz. The output of the nanofibre is then detected on a low noise balanced photoreceiver (see caption of Fig.~\ref{fig:setup}).  The interference between the collected signal field and the local oscillator generates a 72.58~MHz beat note in the photoreceiver output, shifted well away from technical noise sources and amplified compared to direct detection of the signal field  by the factor $2 \sqrt{P_{\rm LO}/P_{\rm sig}} \sim 10^6$~\textendash $~10^8 $, where $P_{\rm LO}$ is the power of the local oscillator field and $P_{\rm sig}$ is the signal power (see supplementary information II).

\subsection{Experimental setup:}
The light from a 780 nm diode laser is split into three beams (see Fig~\ref{fig:setup}b):  
The first beam is frequency shifted using an Acousto Optic Modulator (AOM) and used as probe field. 
This probe field is focused with a microscope objective onto the nanofibre waist. 
The second beam, the local oscillator, goes through the nanofibre. Polarisation controllers are used in both the probe and local oscillator fields to maximise their interferences. After passing through the fibre the combined field is detected on a 
balanced detector 
together with the balance field coming from the last beam of the laser. 
The photocurrent output from the detector is then 
passed through a home-built low electronic noise dual-quadrature lock-in amplifier 
to produce two quadrature signals that are recorded on an oscilloscope. The lock-in amplifier first high-pass filters the photocurrent, then amplifies it, and finally mixes it down using two radio-frequency mixers. The phases of the mixing processes are to generate orthogonal quadrature signals quantifying the envelope amplitude of the sine and cosine components of the photocurrent. Anti-aliasing filters are employed prior to the oscilloscope to prevent mixing of high frequency noise into the recorded signals. A final signal proportional to the scattered field amplitude is generated by taking the quadrature sum of the two quadrature signals.

\subsection{Experimental procedure:}
The nanofibre is positioned on top of a microscope cover slip. With a syringe a $\sim$0.6 mL droplet of ultrapure double processed deionized water (Sigma W3500) is deposited on top of the cover slip and the fibre. The \textit{LOMO} 40$\times$ (0.75 NA) probe objective is immersed into the droplet and focused on the nanofibre waist. Before adding nanoparticles to the water a set of noise calibration data is recorded. 
With a micro pipette, 20 $\mu$L of nanoparticles solution is added to the water at concentration of $1.5 \times 10^8 $, $1.3 \times 10^7 $ and $5.9 \times 10^4 $ particles per mL for 25~nm, 5~nm silica spheres and 10 by 45~nm gold nanorods respectively . 
The nanoparticle concentrations are chosen such that multiparticle events are highly improbable,  between $ 3.0 \times 10^{-7}$ to $ 7.5 \times 10^{-4}$ particles on average within the detection volume. 
Then, continuous time traces with bandwidth of 50--250 kHz are recorded on an oscilloscope (Tektronix MDO3054). Trapping events are identified manually and post processing is performed on a computer (see Supplementary Information VI).

\subsection{Noise floor characterisation:}

In our experimental configuration, the magnitude of the quantum noise introduced by quantisation of light scales differently with  local oscillator power than that of technical noise sources\cite{bachor}. The electronic noise of the detection apparatus is independent of local oscillator power; while the power of the quantum and technical noise introduced by the optical field scale linearly and quadratically, respectively\cite{bachor} . This latter characteristic, which arises due to the introduction of quantum vacuum fluctuations as an optical field is attenuated~\cite{bachor}, provides a rigorous method to characterise whether  the biosensor is quantum noise limited. To perform this characterisation, a spectrum analyser was used to analyse the photoreceiver output at frequencies close to the heterodyne beat frequency under various conditions.

First, we compare the noise power spectrum for the normal operating conditions of the biosensor using an ultrapure water sample containing no nanoparticles or biomolecules, with the electronic noise measured for the same conditions but with  both local oscillator and probe field blocked. 
As shown  in Fig.~\ref{fig:Q_lim}a,  the electronic noise power was found to be a uniform 8.1~dB below the laser noise over a broad frequency band centred on the heterodyne beat frequency.
Spikes are observed in both the electronic and laser noise at the beat frequency. These arise from pick-up in the detection and optical apparatus, respectively and are sufficiently narrow band to be negligible over the frequency band relevant to our measurements.

 Second, to test whether the local oscillator field is quantum noise limited, we characterise the scaling of the measurement noise with the local oscillator power in the absence of the probe field. 
Fig.~\ref{fig:Q_lim}b shows the noise power for a range of local oscillator powers between 0 and 1~mW, averaged over a 10~kHz frequency window centred at the heterodyne beat frequency. 
The observed linear dependence is consistent with a quantum noise model due to quantisation of the local oscillator field, and inconsistent with the power-squared dependence expected for technical noise. We therefore conclude that the local oscillator field is quantum noise limited over the frequency range of interest.

Finally, the measurement can be degraded by probe noise. This noise is introduced along with the collected light intensity when the probe scatters from a trapped particle. To characterise it in isolation from the motion of the particle, we align the probe to a defect on the surface of the nanofibre. This introduces a stationary source of scattering. The total noise power spectrum of the biosensor can then be characterised as a function of the collected signal power, and calibrated to the quantum noise level via measurements using the local oscillator field alone, as discussed in the main text. 

\subsection{Future improvement in precision:}
Several avenues exist that may allow further improvements in precision, without increased risk of photodamage. The noise floor could be reduced by a factor of four by replacing the balanced photoreceiver with a single ultralow noise photodiode, and using homodyne detection and optical phase stabilisation, rather than heterodyne detection. 
Since, in our implementation, the probe intensity was one order of magnitude lower than the local oscillator intensity, further improvements could be obtained by increasing the probe intensity.

\subsection{Nanoparticles and biomolecules:}
The 5 nm and 25 nm radius silica particle are from Polysciences Inc. and the gold nano rods are from Sigma-Aldrich. BSA were purchase from Sigma-Aldrich and Anti- {\it E. coli } antibodies from Australian Biosearch Inc.

\end{methods}

\vspace{5pc}

%

\begin{addendum}
 \item This work was supported by the Australian Research Council Discovery Project (contract no. DP140100734) and by the Air Force Office of Scientific Research and Asian Office of Aerospace Research and Development (grant no. FA2386-14-1-4046). W.P.B. acknowledges support through the Australian Research Council Future Fellowship scheme FF140100650. M.A.T. is supported by a fellowship from the Human Frontiers Science Program.
The authors would also like to thanks Bei-Bei Li for useful discussions. 
 \item [Author contributions] W.P.B.  conceived and led the project. M.A.T. contributed towards the conceptual design.
  N.P.M. performed the experiments and the data analysis, with contributions from L.M.. Samples were prepared by  N.P.M. and M.W. The manuscript was written by  N.P.M., W.P.B., and L.M.
 \item[Competing Interests] The authors declare that they have no
competing financial interests.
 \item[Correspondence] Correspondence and requests for materials
should be addressed to:\\
 Warwick Bowen (email: w.bowen@uq.edu.au).

\end{addendum}

\newpage

{\Large \bf Supplementary Information: Evanescent single-molecule biosensing with quantum limited precision
}

N. P. Mauranyapin, L. S. Madsen, M. A. Taylor, M. Waleed \& W. P. Bowen

\tableofcontents

\newpage

\section{Introduction}
In this supplementary information, we develop a theoretical model  based on Rayleigh scattering in order to find the quantum limit of our sensor. We describe how the experimental data is analysed to obtain the results in the main text and present additional data and simulations supporting our conclusions, including measurements of the motional power spectral density of the trapped particles, measurements of the trap potential as a function of optical local oscillator power, and simulations of the magnitude of the optical component to the trap potential. We discuss control experiments testing for the presence of contaminant particulates, and analysis that allows us to rule out with confidence the presence of aggregation. We finish by comparing our results to the state-of-the-art and explain how we fabricate the nanofibres.

\section{Quantum noise limited measurement of dipole-scattered light}
\label{quantum_noise_limited_measurement_of_dipole_scattered_light}

\subsection{Dipole scattering}
\label{dipole_scattering}

If an optical field containing $n_{in}$ photons is incident on a dielectric spherical particle, in the approximation that the radius $r$ of the particle is much smaller than the optical wavelength $\lambda$ the optical field can -- generally, and certainly for the results presented in the main article -- be well approximated as spatially uniform across the particle, and the scattered field can be well described by dipole scattering. 
The number of photons scattered by the particle is then \cite{taylor_bowen}:
\begin{equation}
n_{scat} \equiv \frac{\sigma}{4 \pi w^2} n_{in} = \frac{(k r)^6}{6 \pi^2}  \left ( \frac{m^2 -1}{m^2+2} \right )^2  \left ( \frac{\lambda}{w} \right )^2 n_{in} \label{nscatt}
\end{equation}
where $w$ is the waist of the optical field, $\sigma$ is the usual dipole scattering cross-section,
%
%
%
$k=2 \pi / \lambda$ is the wave number,  and $ m = n/n_m $ is the ratio of the refractive indices of the particle ($n$) and the medium which surrounds it ($n_m$). We see that, unsurprisingly, the scattered power depends strongly on the radius of the particle, the refractive index contrast, and the strength of focussing of the gaussian beam. The dimensionless ratio
\begin{equation}
\frac{\sigma}{4 \pi \lambda^2} = \frac{(k r)^6}{6 \pi^2}  \left ( \frac{m^2 -1}{m^2+2} \right )^2 \label{dim_rat}
\end{equation}
contains all of the particle dependence in Eq.~(\ref{nscatt}) and defines how strongly the particle will interact with a general electromagnetic field. It therefore quantifies, in an experimental apparatus independent way, the comparative ease with which particles may be detected (the larger the ratio is, the greater the scattering, and the easier the particle will be to detect). We refer to it here as the {\it detectability} of the particle. It is natural to separate the detectability into two parts: a geometrical size factor ${(k r)^6}/({6 \pi^2})$, and the refractive index contrast $({m^2 -1})^2/({m^2+2})^2$. 


\subsection{Quantum noise limit for direct detection of the scattered field}
\label{subsec:Q_lim}

If this scattered field is collected with a collection efficiency $\eta$ and is directly detected, then from Eq.~(\ref{nscatt}) the average photon number observed by the measurement is
\begin{equation}
	\langle n_{det} \rangle =   \frac{\eta}{6 \pi^2}  (k r)^6 \left ( \frac{m^2 -1}{m^2+2} \right )^2  \left ( \frac{\lambda}{w} \right )^2 \langle n_{in} \rangle
\end{equation}
To be confident of the presence of the particle, it must be statistically possible to distinguish this signal from the signal that exists if no particle is present. Assuming that the incident field is shot noise limited, such that the photons incident on the particle are uncorrelated with each other, and the scattering process is linear, the noise if the measurement is dictated by Poissonian statistics, with the variance $V(n_{det}) \equiv \langle n_{det}^2 \rangle -\langle n_{det} \rangle ^2$ of the measurement equal to the mean value, that is $V(n_{det})=\langle n_{det} \rangle$. In this case, the signal-to-noise ratio (SNR) of the measurement is
\begin{equation}
{\rm SNR} = \frac{\langle n_{det} \rangle^2}{V_0(n_{det}) + V(n_{det})}, \label{SNR1}
\end{equation}
where the noise on the denominator has two components: $V(n_{det})$, the variance of the photon number in the presence of a scattering particle, and $V_0(n_{det})$, the variance when no scattering particle is present. Here, since $V(n_{det})=\langle n_{det} \rangle$, we see that, unsurprisingly, $V_0(n_{det})=0$ (we will find later that this is not the case for heterodyne measurement). The signal-to-noise ratio is then 
\begin{equation}
{\rm SNR} =\langle n_{det} \rangle = {\frac{\eta \langle n_{in} \rangle}{6 \pi^2}}  (k r)^6 \left ( \frac{m^2 -1}{m^2+2} \right )^2  \left ( \frac{\lambda}{w} \right )^2  \label{SNR}
\end{equation}

The quantum noise limit of the measurement is then found by setting the signal to noise ${\rm SNR}=1$, resulting in a minimum detectable scattering cross-section
\begin{equation}
\sigma_{min}= \frac{4 \pi w^2}{\eta n_{in}}, \label{min_sig}
\end{equation}
and a minimum detectable particle radius $r_{min}$ of
%
%
\begin{equation}
r_{min} = \frac{1}{2} \left ( \frac{w \lambda^2}{\pi^2}  \right )^{1/3}  \left ( \frac{6}{\eta \langle n_{in} \rangle} \right)^{1/6}  \left ( \frac{m^2+2}{m^2 -1} \right )^{1/3}.\label{min_r}
\end{equation}
As one might expect, improved refractive index contrast (increased $m$), improved collection efficiency, and increased input photon flux ($n_{in}$) all improve the minimum detectable radius. 

Comparison of Eq.~(\ref{dim_rat}) with Eq.~(\ref{min_sig}) shows the significance of the dimensionless ratio defined in Eq.~(\ref{dim_rat}). If this ratio equals one for a given particle, then, at the quantum noise limit, one incident photon would be sufficient to detect the particle with a perfect efficiency detector if the incident field is focussed so that its waist size equals the optical wavelength ($w=\lambda$).

\subsection{Quantum noise limit for heterodyne detection}
\label{q_het_sec}

In heterodyne detection, rather than directly detecting the scattered field, it is instead interfered with a bright local oscillator field, separated in frequency from the incident field by a frequency difference $\Delta$. In a quantum mechanical description the signal field (here, the scattered field) is treated quantum mechanically, while the local oscillator is treated classically with its fluctuations neglected. This is valid so long as the local oscillator field is much brighter than the signal field. In our case this approach is particularly appropriate. The field scattered into the tapered optical fibre has photon flux in the range of $4 \times 10^{3} $ per second, while the photon flux of the local oscillator was $4 \times 10^{15} $ per second, 12 orders of magnitude greater. The combined field can then be expressed in a rotating frame at the frequency of the scattered field as
\begin{equation}
\hat a = \sqrt{N_{\rm LO}}e^{i \Delta t} + \sqrt{\langle N_{det} \rangle} + \delta \hat a, \label{a1}
\end{equation}
where $N_{\rm LO}$ and $N_{det}$ are the  photon flux in the local oscillator and scattered fields reaching the detector, respectively, in units of photons per second, formally $\hat a$ is the annihilation operator of the combined field but informally it may be thought of as an appropriately normalised complex phasor describing the amplitude and phase of the field in phase space, and $\delta \hat a$ is a fluctuation operator with zero mean ($\langle \delta \hat a \rangle = 0$) that includes all of the fluctuations of the scattered field (the shot noise). In quantum mechanics the non-commutation of the annihilation and creation operators ($[\hat a, \hat a^\dagger] \equiv \hat a \hat a^\dagger - \hat a^\dagger \hat a = \delta (t)$, where $\delta(t)$ is the Dirac delta function) is ultimately responsible for the noise floor of the measurement, once all technical noise sources are removed, and therefore for the shot noise.

The combined field is then detected, resulting in a photocurrent proportional to the photon number in the field
\begin{eqnarray}
i &=& \hat a^\dagger \hat a\\
 &=& N_{\rm LO} + \langle N_{det} \rangle + 2 \sqrt{N_{\rm LO} \langle N_{det} \rangle} \cos \Delta t  +  \sqrt{N_{\rm LO}}  \left ( \delta \hat a^\dagger e^{i \Delta t} + \delta \hat a e^{-i \Delta t} \right ) +  \delta \hat a^\dagger \delta \hat a\\
 &\approx& N_{\rm LO} + \langle N_{det} \rangle + 2 \sqrt{N_{\rm LO} \langle N_{det} \rangle} \cos \Delta t  +  \sqrt{N_{\rm LO}}  \left ( \hat X \cos \Delta t + \hat P \sin \Delta t \right ) \label{photo}
\end{eqnarray}
where in the approximation we have made the usual approximation that the product of fluctuations $\delta \hat a^\dagger \delta \hat a$ is much smaller than the other terms in the expression and that the scattered photon flux is much smaller than the local oscillator flux $(N_{LO} >> N_{det}) $. For measurements that approach the quantum noise limit this is appropriate so long as the local oscillator photon number is much larger than one, which is clearly the case here. The quadrature operators $\hat X$ and $\hat P$ are defined as usual as
\begin{eqnarray}
\hat X &=& \hat a^\dagger + \hat a\\
\hat P &=& i \left (\hat a^\dagger -\hat a \right),
\end{eqnarray}
and their commutation relation $[\hat X, \hat P] = 2 i \delta(t)$ results in the Heisenberg uncertainty principle $V(\hat X(t)) V(\hat P(t)) \ge 1$, with the shot noise (or quantum noise) limit of a measurement reached when $V(\hat X) = V(\hat P) = 1$.

It is clear from Eq.~(\ref{photo}) that signals due to the mean scattered photon number $n_{det}$ are present both at zero frequency and in a beat at frequency $\Delta$. Since, as discussed above for heterodyne detection, the local oscillator must be much brighter than the signal field, the zero frequency term is obscured by the the presence of the local oscillator. Furthermore, since the beat term includes the square-root of the local oscillator photon number, its amplitude is much greater than that of the zero frequency term.  Consequently, in heterodyne detection, and in our case to detect the presence of a particle,  the component at frequency $\Delta$ is utilised. To extract this component in our experiments, since the phase of the beat is in practise unknown, we mix the photocurrent down with two electronic local oscillators at frequency $\Delta$ but $\pi/2$ out of phase, and integrate for a time $\tau$. From these two photocurrents, it is possible to extract the magnitude of the beat. The result is equivalent to mixing the photocurrent of Eq.~(\ref{photo}) down in the following way:
\begin{eqnarray}
\tilde i &=&  \int_{t}^{t+\tau} dt \,\,  i \times \cos \Delta t\\
&\approx&  \sqrt{N_{\rm LO}}  \left [ \sqrt{\langle N_{det} \rangle} \, \tau +   \int_t^{t+\tau} \left ( \hat X \cos^2 \Delta t + \hat P \sin \Delta t \cos \Delta t \right ) dt  \right ], \label{itemp}
\end{eqnarray}
where we assume that $\tau$ is sufficiently long to remove components in the photocurrent that oscillate at frequencies fast compared with the beat frequency $\Delta$. Assuming that the fluctuation terms $\hat X$ and $\hat P$ are Markovian white noise, as is the case for a shot noise limited field or any field with a white power spectrum over the frequencies of interest, the fluctuation term in the above equation can be re-defined via Ito calculus as a new Markovian fluctuation operator\cite{gardiner}
\begin{equation}
\tilde X \equiv \sqrt{\frac{2}{\tau}}  \int_t^{t+\tau} \left ( \hat X \cos^2 \Delta t + \hat P \sin \Delta t \cos \Delta t \right ) dt,
\end{equation}
normalised such that the variance $V(\tilde X) = 1$. This results because -- in a classical sense -- both the sign and value of $\hat X$ and $\hat P$ are random (they are, classically, random Gaussian variables) as a function of time. The effect of the sinusoidal envelopes modulating each term is then just to modulate the total power of the noise -- unlike a coherent signal, integration in time of a random variable multiplied by a function such as $\sin \Delta t \cos \Delta t$ does not average to zero in the long time limit. 

We then find
\begin{eqnarray}
\tilde i &=&  \sqrt{N_{\rm LO}}  \left ( \sqrt{\langle N_{det} \rangle} \, \tau + \sqrt{\frac{\tau}{2}} \tilde X  \right )\\
&=& \sqrt{n_{\rm LO}}  \left ( \sqrt{\langle n_{det} \rangle}  + \frac{\tilde X}{\sqrt{2}}   \right ),
\end{eqnarray}
where the total photons numbers $n_{\rm LO}$ and $n_{det}$ are given by $n_{\rm LO}=N_{\rm LO} \tau$ and $n_{det} = N_{det} \tau$, respectively. From this signal we wish to distinguish whether there is a scattering particle in the optical field or not, and can -- as we did in the previous section on direct detection -- use the signal-to-noise ratio in Eq.~(\ref{SNR1}). Unlike direct detection, however, here we find that there is noise in the measurement even when $\langle n_{det} \rangle =0$. This exists due to the presence of the local oscillator, which 
amplifies the vacuum noise of the field. 
We then find that for a quantum noise limited field, with $V(\tilde X)=1$,
\begin{equation}
{\rm SNR} = \langle n_{det} \rangle,
\end{equation}
exactly identical to the expression we obtained for direct detection in Eq.~(\ref{SNR}). We therefore find that, in principle, the quantum noise floors of heterodyne detection and direct detection are identical, and both governed by Eqs.~(\ref{min_sig})~and~(\ref{min_r}).

\subsubsection{Quantum noise limit for heterodyne detection with amplitude noise cancellation}

In our experiments we make one modification to the heterodyne detection scheme described above, following the approach used in Refs.~\cite{freudiger}. Often, and in our experiments, classical laser intensity present in the local oscillator can contaminate the measurements and preclude reaching the quantum noise limit. To eliminate this laser noise, we split the laser field used for the local oscillator into two equal power beams on a beam splitter. One of these beams was interfered with the scattered field, while the other bypassed the experiment. The two beams were then detected on a balanced detector, which allowed the classical amplitude noise to be subtracted. 

Mathematically, extending Eq.~(\ref{a1}), we can describe the two fields that arrive at the detector as
\begin{eqnarray}
\hat a &=& (\sqrt{N_{\rm LO}} + X_c/\sqrt{2}) e^{i \Delta t} + \sqrt{\langle N_{det} \rangle} + \delta \hat a\\
\hat a_{-} &=& (\sqrt{N_{\rm LO}}+ X_c/\sqrt{2})e^{i \Delta t} + \delta \hat a_{-},
\end{eqnarray}
where $X_c$ is the classical intensity noise on the laser, which is correlated between the two fields, while $\hat a_-$ is an annihilation operator describing the cancellation field and $\delta a_-$ is the quantum noise on that field which, due to the action of the beam splitter, is uncorrelated to the noise on the local oscillator field.Working through similar mathematics to that above, and assuming that the two fields are perfectly balanced, and therefore that the classical noise is perfectly cancelled, we find that the single-to-noise ratio is quantum noise limited, but degraded by a factor of two compared with pure heterodyne detection. That is
\begin{equation}
{\rm SNR} = \langle n_{det} \rangle/2.
\end{equation}
In this case, the minimum detectable scattering cross-section $\sigma_{min}$ and particle radius $r_{min}$ are degraded to
\begin{equation}
\sigma_{min}= \frac{8 \pi w^2}{\eta n_{in}}, 
\end{equation}
and 
%
%
\begin{equation}
r_{min} = \left ( \frac{w \lambda^2}{\pi^2}  \right )^{1/3}  \left ( \frac{12}{\eta \langle n_{in} \rangle} \right)^{1/6}  \left ( \frac{m^2+2}{m^2 -1} \right )^{1/3}.
\end{equation}
Knowing that $P_{in} = \hbar \omega \langle n_{in} \rangle / \tau $ with $P_{in}$ the input power, $\hbar$ the reduced Plank constant, $ \omega$ the frequency of the  light and $\tau$ the measurement time we have :

\begin{equation}
\sigma_{min}= \frac{8 \hbar \omega}{\eta \tau I_{in} }
\end{equation}
and
\begin{equation}
r_{min} = \left ( \frac{12 \lambda^4 \hbar \omega}{\pi^5 \eta I_{in} \tau} \right)^{1/6}  \left ( \frac{m^2+2}{m^2 -1} \right )^{1/3}
\end{equation}
with $I_{in} = P_{in}/\pi w^2$ the input intensity.

We note that this kind of intensity noise cancellation has proved to be a powerful technique for precision evanescent biosensing, and have been utilised in a range of previous experiments, see for example Ref.\cite{freudiger}.

\subsubsection{Scaling of quantum and classical noise as a function of loss}

It is generally possible to confirm that an experiment is quantum noise limited by varying the efficiency of the measurement and measuring the effect this has on the signal-to-noise ratio. This is due to differences in the effect of loss on quantum and classical noise
due to the introduction of quantum vacuum noise. This vacuum noise contamination is a necessary consequence of the Heisenberg uncertainty principle. Mathematically, inefficiencies can be modelled by the action of a beam splitter, one output of which is lost, and one input of which introduces the vacuum fluctuations\cite{bachor}. 

To see the effect of inefficiency on our experiment, we begin with Eq.~(\ref{a1}), describing the annihilation operator of the field to be detected via heterodyne detection. In that equation, the detected photon flux $N_{det} = \eta N_{scat}$ where $N_{scat}$ is the scattered photon flux and $\eta$ is the detection efficiency. $\delta \hat a$ is the fluctuation operator that describes the noise on the measurement after the inefficiencies in detection. In our initial treatment, we took this to be purely quantum noise. If we include Markovian classical amplitude quadrature noise prior to the introduction of any losses, and vacuum noise entering due to the presence of the loss then $\delta \hat a$ can be expanded as
\begin{equation}
\delta \hat a = \sqrt{\eta} ( \delta \hat a_{prior} + \delta a_c) + \sqrt{1-\eta} \delta a_v,
\end{equation}
where $\delta \hat a_{prior}$, $\delta a_c$, and $\delta a_v$ are, respectively, the annihilation operators describing the quantum fluctuations on the field prior to any losses, the classical noise, and the vacuum noise introduced by the loss. By examining the variance of the amplitude quadrature $\delta X = \delta \hat a^\dagger+\delta \hat a$ we can gain some insight into the difference between quantum and classical noise:
\begin{eqnarray}
V(\delta \hat X) &=& \eta V(\delta \hat X_{prior}) + \eta V(\delta X_c) + (1-\eta) V(\delta X_v)\\
&=&  \eta V(\delta X_c) +1,
\end{eqnarray}
where we have used the property of quantum fluctuations of coherent light and vacuum fields, that $V(\delta \hat X_{prior}) =V(\delta \hat X_v) =1$. We see, therefore, that while the variance of the classical noise is attenuated with increasing loss (decreasing $\eta$), the quantum noise is unchanged at unity by the action of attenuation. This being the case, we can make the substitution $\delta a_q \equiv  \sqrt{\eta} \delta \hat a_{prior} + \sqrt{1-\eta} \delta a_v$, treating the combined quantum noise as one single input quantum vacuum field. Going through the same calculation as in the main part of Section~\ref{q_het_sec} but including the classical noise term by making the substitution $\delta \hat a \rightarrow \delta \hat a + \sqrt{\eta} a_c$ we reach the photocurrent
\begin{eqnarray}
\tilde i &=&  \sqrt{n_{\rm LO}}  \left ( \sqrt{\langle n_{det} \rangle}  + \frac{\tilde X+ \sqrt{\eta} \tilde X_c}{\sqrt{2}}   \right )\\
&=& \sqrt{\eta \, n_{{\rm LO}, prior}}  \left ( \sqrt{\langle n_{det} \rangle}  + \frac{\tilde X+ \sqrt{\eta} \tilde X_c}{\sqrt{2}}   \right )
\end{eqnarray}
where $n_{{\rm LO}, prior}$ is the number of photons used in the local oscillator prior to any losses, and $\tilde X_c$ is defined, for the classical noise, in the same way as $\tilde X$. The variance of the measurement is then
\begin{equation}
V(\tilde i ) = \frac{n_{{\rm LO}, prior}}{2} \left ( \eta +  \eta^2 V(\tilde X_c) \right ).
\end{equation}
We see that the variance of the quantum noise (first term in the brackets) scales linearly with efficiency $\eta$ while the classical noise scales quadratically. Therefore, by quantifying the measurement noise floor as a function of attenuation, as was performed in the main paper, it is possible to unambiguous determine in which regimes classical and quantum noise dominate, and therefore whether the quantum noise limit has been reached.


\subsubsection{Homodyne detection}

An alternative approach to quantum limited measurement is to perform homodyne detection. This is very closely related to heterodyne detection but uses a local oscillator whose frequency matches the scattered (signal) field frequency (i.e. $\Delta =0$). As can be seen from inspection of Eq.~(\ref{itemp}), this choice of frequency has the immediate advantage of eliminating one of the two noise terms contributing to the measurement, and therefore can be used to improve the quantum noise limited measurement signal-to-noise ratio by a factor of two. However, heterodyne detection has the major advantage that the signals of interest are shifted up to sideband frequencies near the beat frequency $\Delta$. In homodyne detection, this is not the case. This has severe consequences for biophysical applications where the signals to be measured typically reside in the hertz-kilohertz frequency range (in our case, typically beneath 100 Hz), a frequency range in which many low frequency technical noise sources reside\cite{taylor_bowen}. For our experiments, such noise sources were found to preclude quantum noise limited operation by many orders of magnitude when using homodyne detection. Homodyne detection has the further disadvantage that it requires the local oscillator and scattered field to be phase locked with high precision. In particle scattering experiments in liquid this is made difficult both by the weakness of the scattered field and by motion of the particle which changes the path length of the scattered field. It was for these reasons that heterodyne detection was chosen for the results reported here.

\section{Quantum noise limit of cavity enhanced measurement}

There have been significant recent efforts to use the enhanced light-matter interactions available in optical microcavities to allow precision nanoparticle and biomolecule detection (see for example Refs.~\cite{dantham}). Here, we derive the quantum noise limit for such measurements. In the usual approach -- reactive microcavity based sensing\cite{arnold_khoshsima} --the action of the nanoparticle or biomolecule is to change the average refractive index within the resonator and therefore shift its resonance frequency. For a cavity that is initially driven with on-resonance light, the dynamics of the field within the cavity, in the presence of a biomolecule or nanoparticle induced frequency shift $\delta \omega$, is (see for example Ref.~\cite{BowenBook})
\begin{equation}
\dot{\hat a} = - (\kappa/2 + i \delta \omega) \hat a + \sqrt{\kappa} \hat a_{in},
\end{equation}
where $\kappa$ is the cavity decay rate, $\hat a_{in}$ is the annihilation operator describing the incident field, and we assume -- to obtain a bound for the best possible predicted sensitivity -- that there is no loss within the cavity, other than back through the input coupler. In the realistic regime where the particle can be treated as stationary over the characteristic timescales of the cavity dynamics, this equation can be solved by taking the steady-state solution where $\dot{\hat a}=0$. We then find that
\begin{equation}
\hat a = \frac{\sqrt{\kappa}}{\kappa/2 + i \delta \omega} \hat a_{in}.
\end{equation}
The input-output relation\cite{BowenBook} $\hat a_{out} = \hat a_{in} - \sqrt{\kappa} \hat a$ can then be used to determine the out-coupled field:
\begin{equation}
\hat a_{out} = - \left ( \frac{\kappa/2 - i \delta \omega}{\kappa/2 + i \delta \omega}  \right ) \hat a _{in}.
\end{equation}
As before, the input field can be expanded as a bright coherent classical field $\langle a_{in} \rangle = \sqrt{\langle N_{in} \rangle}$ and a fluctuation term $\delta a_{in}$ with $\langle \delta a_{in} \rangle=0$, with $\langle N_{in} \rangle$ being the mean photon flux incident on the cavity. We then obtain
\begin{eqnarray}
\hat a_{out} &=& - \left ( \frac{\kappa/2 - i \delta \omega}{\kappa/2 + i \delta \omega}  \right ) \left ( \sqrt{\langle N_{in} \rangle} + \delta \hat a_{in} \right )\\
&=&- \left ( \frac{\kappa^2/4 - i \kappa \delta \omega - \delta \omega^2}{\kappa^2/4 +  \delta \omega^2}  \right ) \left ( \sqrt{\langle N_{in} \rangle} + \delta \hat a_{in} \right ).
\end{eqnarray}
Assuming that the frequency shift is small compared to the cavity decay rate, and the optical fluctuations are much smaller than $\sqrt{ \langle N_{in} \rangle} $ we can neglect all terms in this expression that include either $\delta \omega^2$ or the product $\delta \omega \delta a_{in}$, with the result
\begin{eqnarray}
\hat a_{out} 
&=&- \left ( 1 - \frac{4 i \delta \omega}{\kappa}  \right )  \sqrt{\langle N_{in} \rangle} - \delta \hat a_{in}.
\end{eqnarray}
It is apparent from this expression that the first order effect of the particle on the output optical field is to shift its phase. The phase quadrature of the output field is
\begin{equation}
\hat P_{out} = i \left ( \hat a_{out}^\dagger - \hat a_{out} \right ) =  \frac{8 \delta \omega}{\kappa}   \sqrt{\langle N_{in} \rangle} - \delta \hat P_{in}.
\end{equation}
A homodyne measurement can detect the phase quadrature, in principle, without any additional noise, resulting in a photocurrent integrated over the time $\tau$ of
\begin{equation}
i \propto \int_t^{t+\tau} dt \, \hat P_{out} = \frac{8 \tau \delta \omega}{\kappa}   \sqrt{\langle N_{in} \rangle} - \int_t^{t+\tau} dt \,  \delta \hat P_{in}.
\end{equation}
Similar to our previous treatment of the quantum noise limit for heterodyne detection, from Ito calculus\cite{wiseman} for Markovian fluctuations (such as the quantum noise of a coherent laser)  the integral $\int_t^{t+\tau} dt \,  \delta \hat P_{in} = \sqrt{\tau} \delta \tilde P_{in}$, where $\delta \tilde P_{in}$ is a Markovian noise process with -- for a shot noise limit field -- variance $V(\delta \tilde P_{in})=1$. The detected photocurrent is then
\begin{equation}
i \propto \frac{8 \delta \omega}{\kappa}   \sqrt{\langle n_{in} \rangle} -  \delta \tilde P_{in},
\end{equation}
where as before $n_{in} = \tau N_{in}$ is the total photon number incident on the detector during the measurement time. Using Eq.~(\ref{SNR}) and assuming the incident field is shot noise limited, the signal-to-noise ratio for discrimination of the presence of the frequency shift due to the particle is then
\begin{equation}
{\rm SNR} = \frac{({8 \delta \omega}/{\kappa})^2 n_{in}}{2 V(\delta \tilde P_{in})} =  32  \left (\frac{\delta \omega}{\kappa} \right )^2 n_{in}.
\end{equation}
Setting ${\rm SNR}=1$, we find the minimum detectable frequency shift
\begin{equation}
\delta \omega_{min} = \frac{\kappa}{\sqrt{32 \, n_{in}}}. \label{min_freq_shift}
\end{equation}

It now remains to determine the frequency shift introduced by a particle within the optical field.
In first order perturbation theory, assuming that the particle is located at the position of peak intensity within the optical mode, the frequency shift it induces is given by \cite{arnold_khoshsima}
\begin{equation}
\delta \omega = - \frac{\alpha \Omega}{2 V},
\end{equation}
where $\alpha$ is the polarizability of the particle, $\Omega$ is the bare cavity frequency, and
\begin{equation}
V \equiv \frac{\int \epsilon_r |{\bf E} ({\bf r})|^2 d {\bf V}}{{\rm max} \{ |{\bf E}({\bf r})|^2 \}}
\end{equation}
is the mode volume of the cavity optical eigenmode, with ${\bf E} ({\bf r})$ being the electric field distribution of the mode,  $\bf r$ being a spatial co-ordinate in three dimensions, and $\epsilon_r$ being the relative permittivity of the cavity medium.

For the case of a dielectric sphere, at optical wavelengths the polarizability is~\cite{jackson}
\begin{equation}
\alpha = 4 \pi \epsilon_m r^3 \left ( \frac{m^2 - 1}{m^2+2} \right ),
\end{equation}
where $\epsilon_m = n_m^2$ is the relative permittivity (or refractive index square) of the surrounding medium. For a sphere, the particle-induced optical frequency shift is therefore
\begin{equation}
\delta \omega = - \frac{2 \pi \epsilon_m r^3}{V} \left ( \frac{m^2 - 1}{m^2+2} \right ) \Omega.
\end{equation}
Substituting this expression into Eq.~(\ref{min_freq_shift}) for the minimum detectable frequency shift and rearranging, we finally arrive at the minimum detectable scattering cross section and radius using cavity enhanced sensing
\begin{eqnarray}
 \sigma_{min} &=& \frac{ k^4 V^2}{48 \pi \epsilon_m \, Q^2 \, n_{in}}\\
 r_{min} &=& \left ( \frac{V}{2 \pi \epsilon_m Q }\right )^{1/3} \left ( \frac{m^2+2}{m^2 - 1} \right )^{1/3}  \left ( \frac{1}{32 \, n_{in}} \right )^{1/6},
\end{eqnarray}
where we have defined the optical quality factor $Q \equiv \Omega/\kappa$.

\section{Quantum noise limit of our nanofibre sensor}
\label{num_app}

In this section we quantify the best sensitivity achievable by cavity enhanced and heterodyne detection sensors and compare these predictions with examples of the state-of-the-art:

For typical parameters used in our heterodyne nanofibre experiments, with an input probe power of 
$ P_{in} = 2 \pi c \hbar n_{in} / \lambda \tau = 2$~mW, with $ \hbar $ the Planck constant, c the speed of light, $ \lambda = 780$~nm and 
$ \tau=0.01 $~sec the measurement time,
a probe beam waist of $w = 3 \mu$m, 
$ n = 1.45 $, $ n_m = 1.33 $ and
a collection efficiency $\eta = 0.01$,  
we find the theoretical minimal cross section detectable to be $ \sigma_{min_{fiber}} = 3 \times 10^{-4} $~nm$^2$ equivalent to a silica sphere radius of $ r_{min_{fiber}} = 23 $~nm.

For the same input power and for typical microcavities with  
a quality factor of $ Q = 3 \times 10^8 $,
$ n_m = \sqrt{ \epsilon_m } = 1.33   $, 
and a mode volume $ V = 350 \mu $~m~$^2$,
we find that the minimal cross section detectable is $ \sigma_{min} = 1.7 \times 10^{-7} $~nm$^2  $ corresponding to a silica particle with a radius of $ r_{min} = 3.0 $nm.

These theoretical predictions show that quantum noise limited microcavities should be able to detect particles with cross section three orders of magnitude smaller than nanofibre sensors, corresponding to a reduction in radius for a silica particle of a factor of eight. Unfortunately to date it has proved difficult to reach this limit with such biosensors and their sensitivity is at the same order of magnitude as heterodyne detection (see section \ref{state_of_the_art}).


\section{Power spectral density of trapped particles}
\label{tf_np}

\begin{figure}
\centering
\includegraphics[width=1\textwidth,height=0.6\textwidth,clip,keepaspectratio]{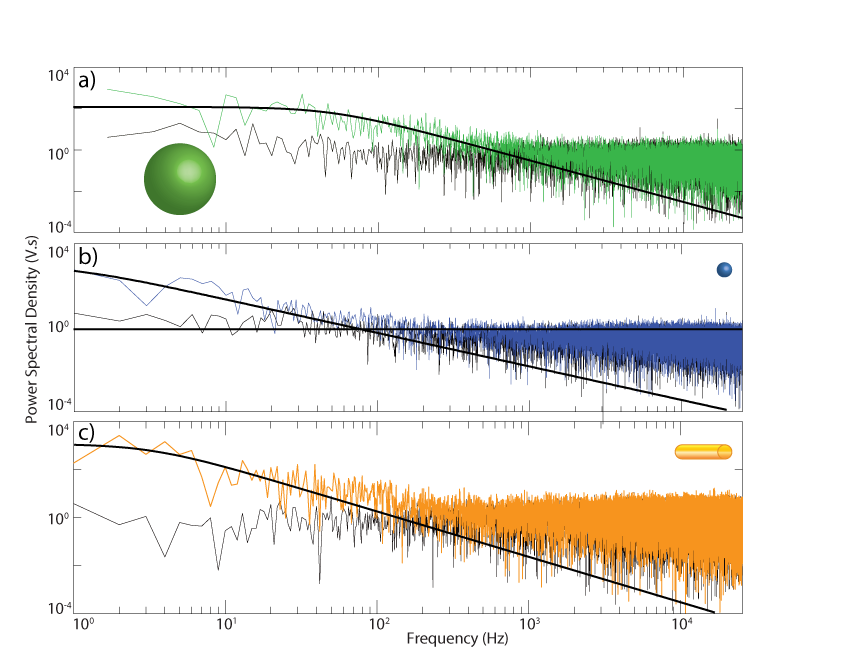}
\caption{ Fourier transform of the trapping events of the figure 3a,b,c (green for 25 nm silica particle, blue for 5 nm silica particle and orange for 5 by 45 nm gold nano rod). The black traces represent the laser noise of each experiment and the color traces are fitted with a Lorentzian shown in black line.
}
\label{fig:TF_NP1}
\end{figure}

\begin{figure}
\centering
\includegraphics[width=1\textwidth,height=0.6\textwidth,clip,keepaspectratio]{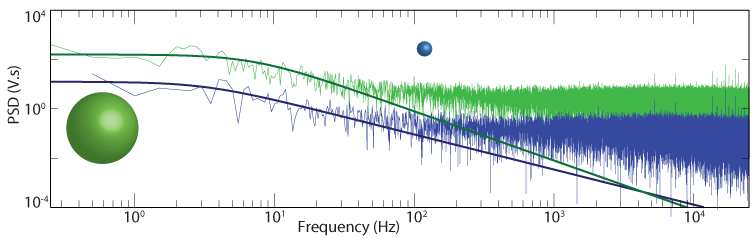}
\caption{ Fourier transform of the trapping events of the figure 3d for a combined solution of 25 and 5 nm silica particle
}
\label{fig:TF_NP2}
\end{figure}

\begin{figure}
\centering
\includegraphics[width=1\textwidth,height=0.6\textwidth,clip,keepaspectratio]{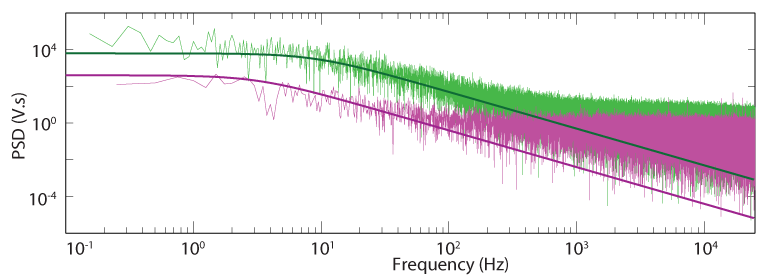}
\caption{ Fourier transform of the trapping events of the figure 4 for a combined solution of BSA (pink) and antibody (green).
}
\label{fig:TF_bio}
\end{figure}

As shown in Ref. \cite{Berg} the power spectral density of a trapped particle can be approximated by a Lorentzian function and its corner frequency indicates the stiffness of the trap. Figures \ref{fig:TF_NP1}, \ref{fig:TF_NP2},\ref{fig:TF_bio} show the power spectral density of each trapping event displayed in the main text, confirming that the particles are trapped by our system.

\section{Probability distribution and trapping potential calculations}
\label{pot_distrib_calculation}

When a particle is scattering probe light close to the nanofibre more light will be collected than when the particle is further from the fibre, thereby modulating the amplitude of the recorded signal field. 
This relation between the particle position and the amplitude of the signal has the same shape as the optical field and decays approximately exponentially following the relation $r = -k^{-1}log(\langle n_{det} \rangle)$ where $r$ is the position of the particle relative to the fibre , $k=2\pi n_{m}/\lambda$ is the wave number, and $\lambda = 780nm$ is the wavelength of the light~\cite{swaim}. 
The position of the nanoparticle can then be calculated from the amplitude relative to the position with the highest amplitude.  
A histogram of the position generates the position probability density $p_{density}(r)$ for each event. According to Boltzmann statistics it is related to the potential $U(r)$ experienced by the trapped particle by $U(r)/k_BT = - log(p_{density}(r))$.


\section{Precision}

To calibrate our sensor we can calculate the resolution with which we can track the radial position of the particle. From the section \ref{pot_distrib_calculation}, we have the relation between the detected signal $ \langle n_{det} \rangle $ and the radial position $r$. The error in position variation is then:
\begin{equation}
	\delta r = - \frac{2 \pi n_m}{\lambda} log( \langle n_{det} \rangle - \delta n_{det} ) + \frac{2 \pi n_m}{\lambda} log( \langle n_{det} \rangle + \delta n_{det} ) 
\end{equation}
\begin{equation}
	\delta r = \frac{2 \pi n_m}{\lambda} log \left( \frac{\langle n_{det} \rangle + \delta n_{det} }{ \langle n_{det} \rangle - \delta n_{det}} \right) 
\end{equation}
with $ \delta n_{det} $ the standard deviation of the detected signal noise.

From the results displayed in figure 3 of the main text, we find that for the 5 nm and 25 nm silica nanoparticle $\langle n_{det} \rangle /  \delta n_{det} $ is equal to 37 and 138 respectively with average over 10 ms. The resolution is then 5 nm and 1 nm respectively with a 100 Hz bandwidth.

\section{Attractive forces}
\label{attractive_forces}

Because the power spectral density shape of the nanoparticles detection event is Lorentzian (see section \ref{tf_np}), we have strong evidences that they are trapped. In conventional nanofibre traps, a combination of attractive optical gradient force and repulsive electrostatic forces are used together to trap the particle next to the fibre \cite{swaim}. Similar to optical tweezers, particles diffusing in the evanescent field around the fibre will be polarised and then attracted toward the centre of the fibre following the gradient of the light intensity. In our case, we theoretically (see following subsection \ref{modelling_pot}) and experimentally (see following subsection \ref{influence_lo}) show that this optical force is not powerful enough explain the trapping of nanoparticles observed in our experiment.
Alternatively, the attractive force could, in principle be introduced by gravity, as observed in some total internal reflection microscopy (TIRM) experiments \cite{gong}. However, we found that these forces are roughly five orders of magnitude too small to explain the attractive forces observed here (see section \ref{gravity}).

\subsection{Modelling of the optical trapping potentials}
\label{modelling_pot}

\begin{figure}
\centering
\includegraphics[width=1\textwidth,height=0.6\textwidth,clip,keepaspectratio]{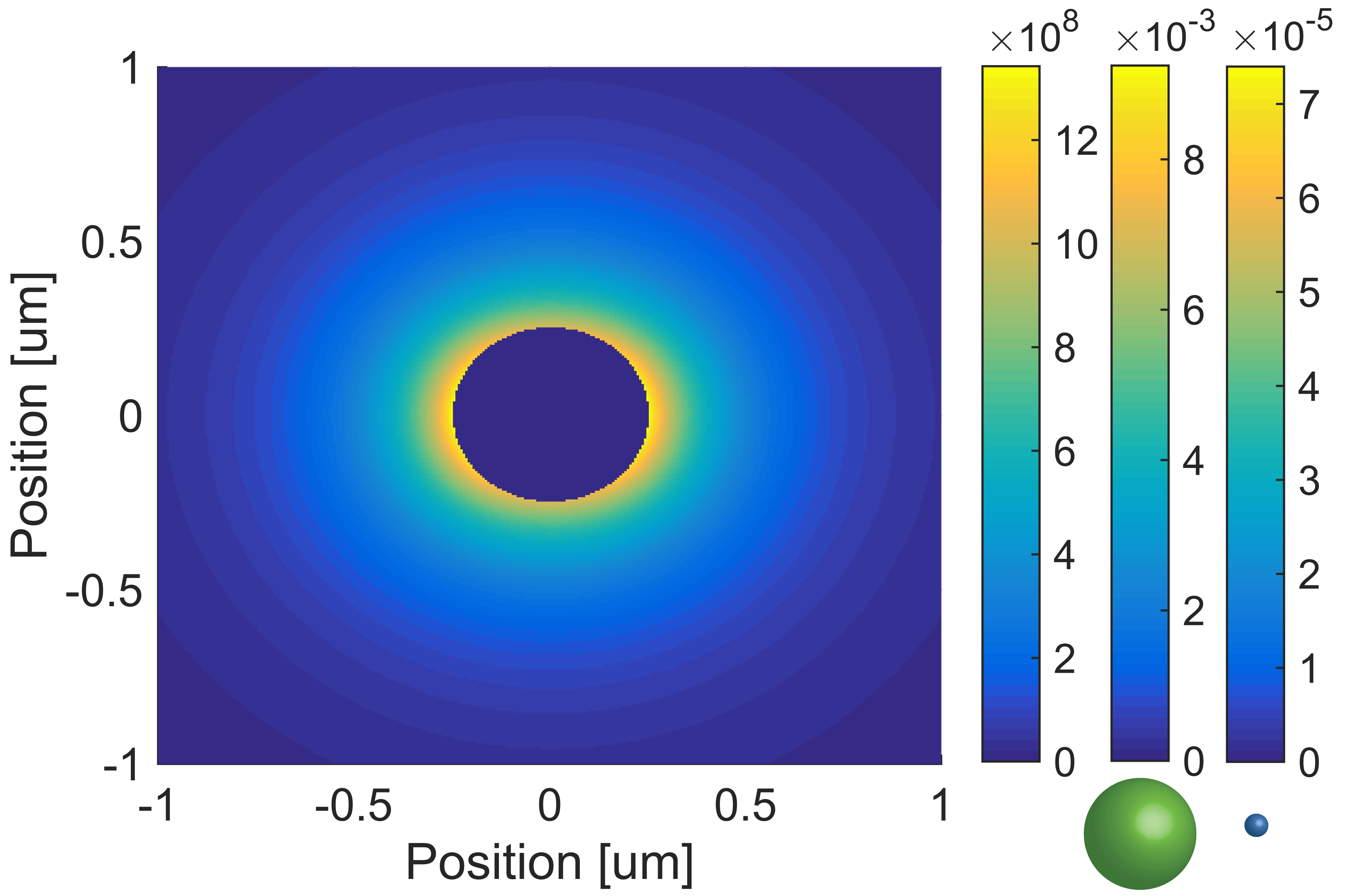}
\caption{Evanescent field intensity map around the nanofibre. As a visual aid the intensity inside the fibre is set to zero giving the centered disk. The first colorbar shows the intensity in units of ${W}/{m^2}$. At a distance of 100 nm from the fibre surface where the particle are expected to be trapped, the intensity is $8 \times 10^8W/m^2$. The second and third colorbars are the trapping potentials normalised to $-k_B T$ for the 25 nm and 5 nm particles respectively as marked with the particles (not drawn to scale with the fibre). Note that in both cases the particles should not be trapped as the potential is below $ k_BT$.
}
\label{fig:intensity}
\end{figure}

In a step index optical fibre the optical field is guided by total internal reflection caused by the refractive index difference between the core and the cladding. Optical nanofibres works in the same way with the core of silica and the cladding made of the surrounding medium. As the light is guided by total internal reflection an evanescent optical field extends out of the fibre. Theoretical models for the extend and intensity of the evanescent fields are well developed in \cite{snyder,kien,vetsch}. The electric fields can be found analytically using the model for a step-index fibre, however finding the propagation constant is a numerical task. Here we follow reference \cite{vetsch} closely and find the propagation constant, get the normalisation for the electric fields, and calculate electric fields on a 500 by 500 grid in a 4 $\mu m^2$ area centered at the nanofibre. From the electric field we calculate the intensity as shown in Fig. \ref{fig:intensity} for 1 mW of horizontally polarised 780 nm light. The trapping potential $U(r)$ is obtained as function of the particle radius $r$ and polarisability $\alpha(r)$ as
\begin{equation}
\begin{split}
U(r)&= -\frac{1}{4} \alpha(r) |\vec{E}|^2, \text{with} \\
\alpha(r)&=4 \pi \epsilon_0 n_m^2 r^3 \frac{m^2-1}{m^2+2} , 
\end{split}
\end{equation}
where $n_m=1.33$ is the refractive index of water.

\begin{figure}
\centering
\includegraphics[width=1\textwidth,height=0.6\textwidth,clip,keepaspectratio]{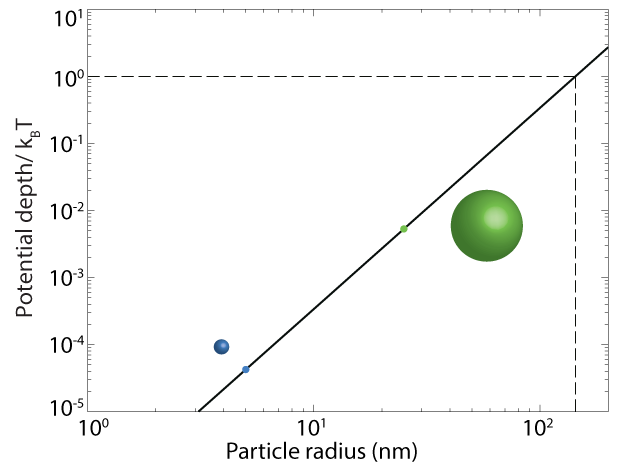}
\caption{ Theoretical trapping potential depth in function of the silica particle radius 100~nm away from the fibre surface. The blue and green points represent the 5~nm and 25~nm silica particles trapped in our experiment, respectively.
}
\label{fig:depth_vs_size}
\end{figure}

To trap particles, the depth of the nanofibre potential must be at least equal to the thermal energy $k_BT $. Our modelling shows that this only occur for silica nanospheres of radius above 145 nm with our experimental parameter (1 mW optical local oscillator power). The 25 nm and 5 nm particles we trap are predicted to have an optical potential depth of, 3 and 5 orders of magnitude smaller than $k_BT$, respectively (see Figs. \ref{fig:intensity} and \ref{fig:depth_vs_size}).

\subsection{Influence of local oscillator on attractive forces}
\label{influence_lo}

\begin{figure}
\centering
\includegraphics[width=1\textwidth,height=0.6\textwidth,clip,keepaspectratio]{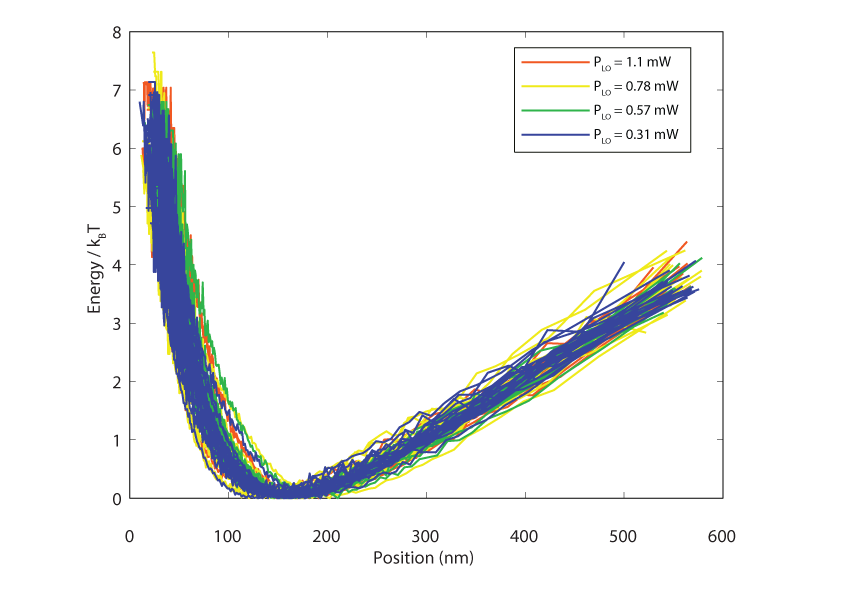}
\caption{ Trapping potential of all observed trapping event of 25 nm silica particles for different L.O. power: red 1.1 mW, yellow 0.78 mW, green 0.57 mW and blue 0.31 mW.
}
\label{fig:optical_trap}
\end{figure}

To verify experimentally that the optical forces are not responsible for the attractive trapping forces we vary the optical local oscillator power and examine its influence on the trapping potential. In Fig. \ref{fig:optical_trap} the local oscillator power is varied from 1.1 mW to 0.31 mW, which if optical forces are playing a role should modify the potential. However no significant changes in the trapping potential are observed in agreement with the theoretical prediction that the optical attractive force is negligible (see Fig.~\ref{fig:optical_trap}).

An explanation can be that this attraction comes from long range electrostatic forces. Those forces come from a deformation of the counter-ion layer or electric double layer surrounding a charged particle in solution \cite{nagornyak,xu,larsen,bowen,krishnan}. This layer becomes larger as the ion concentration decreases and can be expected to be significant in deionized water. In the main text, we show the influence of the electric double layer on the biosensor by changing the ion concentration using salt water.

\subsection{Effect of the gravitational forces}
\label{gravity}

As used in total reflection microscopy, the gravitational forces could explain the attractive part of our potential if the particles were trapped on top of the fibre. The gravitational potential is given by \cite{banerjee}:
\begin{equation}
	U_{grav} = \frac{4}{3} \pi r^3 \Delta \rho gh
\end{equation}
with $ r $ the radius of the nanoparticle, $\Delta \rho$ the density difference between the nanoparticle and its surrounding medium, $g$ the gravity acceleration and $h$ the hight. For our silica nanoparticles, $r=25$~nm,  $\Delta \rho = 1650$~kgm$^{-3}$ which gives a normalised potential of $ U_{grav}/hk_BT = 2.6 \times 10^{-7} $~nm$^{-1}$ compare to the $4.3 \times 10^{-2}$~nm$^{-1}$ observed on the figure 3 of the main text. We then conclude that the gravitational forces are five orders of magnitude smaller than what we observe and can not be attributed to the attractive force of our potential.

\section{Cross section enhancement}
\label{discrepancy}

The quantum noise limit calculations in section \ref{num_app} show that, without some form of scattering cross section enhancement, 5 nm silica particles should not be detectable by our sensor. Moreover, from figure 3 of the main text we observe that there is only a factor of $\sim 4$ between the signal amplitude of the 5~nm and 25~nm silica particles. According to the dipole scattering theory developed in section \ref{dipole_scattering}, the signal amplitude scales as particle size-cubed, therefore, the signal from a single 5~nm particle should be more than two orders of magnitude smaller than that from a 25~nm particle.
 

We are confident that, this signal enhancement is not due to aggregation. The concentration of the particles was chosen to be very low in the detection volume ($ 3.0 \times 10^{-7}$ to $ 7.5 \times 10^{-4}$ particles on average within the detection volume), and the samples were sonicated for at least 15 min before being studied. Furthermore, the measured trapping potentials, probability densities and signal amplitudes are highly reproducible as shown in Fig. 3f of the main text where we display, as shaded bands, the potential standard deviation calculated from 9 and 8 events for the 25 nm and 5 nm nanoparticles respectively. If aggregates were being detected, we would expect these parameters to vary with the size of the aggregate. In addition, the signal enhancement is not due to contamination of the sample by outside particles as extreme care was taken when preparing and cleaning the apparatus. Moreover, control experiments with ultra pure water and salt water, both without nanoparticles were conducted and did not exhibit any events for the entire 45 minutes of the experiments, a duration comparable to the experiments performed in the main text.


{

Our hypothesis to explain this enhancement is due to the formation of an electric double layer created by surface charge of the particle. The electric double layer is known in rheology to affect the hydrodynamic radius of particles. This phenomena has already been studied on silica nanospheres by Dynamic Light Scattering and has an important contribution to the hydrodynamic radius especially when using ultra pure water as the concentration of ion is very low the electric double layer become thicker \cite{killmann}. 
The effect of the double layer on the polarizability of charged particle has also been demonstrated in Ref.\cite{park}. They show that without the polarizability enhancement of the electric double layer the potential of micelles smaller than 200 nm is not deep enough for the particles to be stably trapped by an optical tweezers. However, the contribution to the polarizability from the electric double layer is sufficient to enable trapping.

Surface charge enhancement of the scattering cross section may also explain other observations in the literature. For instance, in Ref.\cite{armani} a range of small biomolecules, including steptadivin, antibodies and Cy5, are detected using a microcavity resonator. Without some enhancement mechanism, these molecules are too small be observable by their apparatus and give a signal several orders of magnitude larger than expected. They attributed this signal enhancement to a thermo-optic effect. However, it was demonstrated in Ref~\cite{arnold_shopova} that this effect is also too weak to explain the observed signal magnitude and should produce signals three orders of magnitude smaller than were observed. The experiments in Ref.\cite{armani} used pure water, so that surface charges could be expected to produce a counter-ion distribution of significant spatial extent and therefore an enhanced scattering cross section.

%


\section{Detected particle state of the art}
\label{state_of_the_art}

The state-of-the-art of nanoparticle and biomolecule detection to-date with evanescent biosensors is presented in Fig. \ref{fig:photodamage} as function of the light intensity used to detect them. The state-of-the-art is compared to results of our experiments (red points). As can be seen, our sensor is still competitive with other sensors but uses four orders of magnitude less intensity, reducing the induced photo damage on the sample.

\begin{figure}
\centering
\includegraphics[width=1\textwidth,height=0.6\textwidth,clip,keepaspectratio]{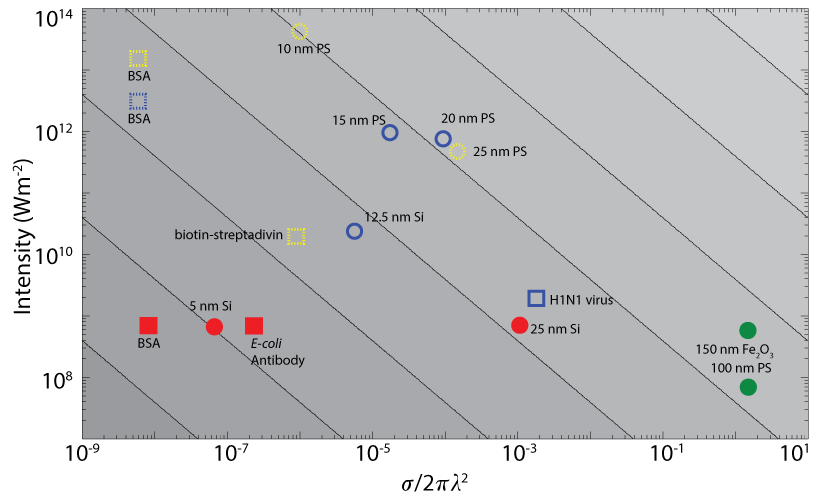}
\caption{\textbf{Overview of particle detection experiments with evanescent biosensors.} Comparison of state-of-the-art experiments. Normalised cross-section of the detected particles versus intensity experienced by the particles.
Detection of nano particles are represented by circles and detection of bio-molecules by squares. The color indicate the techniques used: yellow corresponds to techniques using plasmon resonances, blue to cavity enhanced measurements, dashed blue to hybrid cavity enhanced measurement, green for nanofibre techniques and red represents the particles detected  in the main text. Following the contours means that for a given sensitivity, smaller particles can be detected by increasing the incident intensity. 
Nano-particle detection from left to right: 5 nm silica (Si) [this paper] , 10 nm polystyrene (PS) \cite{plas_10}, 12.5 nm PS \cite{wgm_12.5}, 15 nm PS \cite{wgm_15}, 20 nm PS \cite{wgm_20}, 25 nm PS \cite{plas_25}, 25 nm Si [this paper], 50 nm PS \cite{yu} , and 150 nm \cite{swaim} $Fe_{2}O_{3}$. Bio-molecule detection form left to right: BSA \cite{pang},BSA \cite{dantham}, BSA [this paper], anti \textit{E-coli} antibody [this paper], biotin-streptadivin \cite{zijlstra} and influenza A virus \cite{bio_wgm_virus}.
}
\label{fig:photodamage}
\end{figure}

\section{Tapered fibre fabrication}
Nanofibres are fabricated by pulling a regular 780 HP optical fibre from \textit{Thorlabs Inc.} under a 300 sccm torch of hydrogen. Pulling is stopped when the fibre become single mode again and its diameter reaches about 560 nm~\cite{tong} . Our transmission is between 90 and 99\%.

\newpage

{\Large {\bf References}}
\vspace{2pc}

\bibliographystyle{naturemag}

\bibliography{biblio}

\end{document}